\documentclass{PoS}
\usepackage{url}
\usepackage{wrapfig}
\usepackage{lineno}
\usepackage{wrapfig, blindtext}
\usepackage{subcaption}
\definecolor{gray}{RGB}{200, 200, 200}
\usepackage{soul}

\title{The SPICEcore Hole Camera System}
\ShortTitle{The SPICEcore Hole Camera System}
\author{
The IceCube Collaboration\footnote{For collaboration list, see PoS(ICRC2019)1177} \\
{\href{https://icecube.wisc.edu/collaboration/authors/icrc19_icecube}{https://icecube.wisc.edu/collaboration/authors/icrc19\_icecube}} \\
E-mail: \email{hrvoje.dujmovic@icecube.wisc.edu, minjin.jeong@g.skku.edu, christoph.toennis@gmx.de} 
}
\abstract{
 IceCube is a cubic-kilometer scale neutrino telescope located at the geographic South Pole. The detector utilizes the extremely transparent Antarctic ice as a  medium for detecting Cherenkov radiation from neutrino interactions. As a result of extensive studies of the optical properties of ice, the light propagation in IceCube is well understood. The ice properties are, however, still dominant sources of detector systematic uncertainties in many IceCube analyses. We have designed a camera system to measure the optical properties of the Antarctic ice surrounding the SPICEcore hole that is an ice-core hole drilled down to 1.7~km near the IceCube detector. The device uses CMOS image sensors to measure the back-scattered light from bright LEDs pointing into the ice. Having a similar measurement principle, the device can also serve as a proof of concept of a camera system designed for the optical modules for IceCube Upgrade. During the 2018/2019 austral summer season, a prototype of the instrument was deployed in the ice-core hole. In this contribution, we present the hardware design of the camera system and the result of the first deployment at the South Pole.

\vspace{4mm}
{\bfseries Corresponding authors:} 
H. Dujmovi\'c$^{1}$, M. Jeong$^{1}$, \speaker{C. T\"onnis$^{1}$} \\
{$^{1}$ \itshape Department of Physics, Sungkyunkwan University, Suwon 16419,  Korea}\\
}

\FullConference{36th International Cosmic Ray Conference -ICRC2019-\\
        July 24th - August 1st, 2019\\
        Madison, WI, USA}

\begin{document}

\section{Introduction}
The IceCube Neutrino Observatory~\cite{detector} is the world-largest neutrino telescope located at the geographic South Pole. The observatory consists of one cubic kilometer-scale Cherenkov radiation detector deployed in ice at depths between 1,450~m and 2,450~m, as well as a square kilometer large cosmic ray air shower detector on the surface of the ice. The main scientific goals of the experiment are to measure the high-energy astrophysical neutrino flux and to identify its sources. In 2013, the IceCube collaboration reported the observation of astrophysical neutrinos~\cite{HESE2013}, and in 2018, the collaboration  identified a flaring Blazar as a source of the astrophysical neutrinos~\cite{TXS}. Other scientific goals of the detector include measurements of neutrino oscillation parameters, cosmic ray flux, and the search for dark matter and other exotic particles. \\
\indent For the reconstruction of neutrino events in the IceCube detector, an accurate understanding of the optical properties of the Antarctic ice is crucial. To achieve this, various efforts were made, in particular, using the LED flasher system.  From these efforts, a model of the optical properties of the ice was constructed~\cite{SPICE}. The optical properties of the ice, however, still remain a dominant detector systematic uncertainty in many IceCube analyses. \\
\indent The SKKU SPICEcore hole camera system is an instrument that can measure optical properties of the Antarctic ice, like the scattering length of light in ice, in the South Pole Ice core hole (SPICEcore hole). The system contains CMOS image sensors and light sources, and can measure the light back-scattered by the ice surrounding the bore hole for different depths and orientations. The details of the measurement principle are explained in Section~\ref{method}. The core objective of the camera system is to make measurements that complement the current understanding of the ice. Additionally, these measurements can prove the feasibility of using cameras for calibration purposes as envisioned for the IceCube Upgrade~\cite{ICU_camera}. Data collected with the SPICEcore hole camera system can aid in the development of analysis tools for the IceCube Upgrade camera calibration measurements.  

\section{The Measurement}
\subsection{The SPICEcore hole}

The SPICEcore hole was drilled during the 2014/2015 and 2015/2016 austral summer seasons  near the Amundsen-Scott South Pole Station for the purpose of glaciological, geological and climatological studies~\cite{SPICEcore_hole}. The hole is located about 1~km away from the IceCube detector site and has a depth of 1,751~m, which partially overlaps in depth range with the IceCube detector. IceCube detector modules are buried below 1,450~m, but due to favorable ice tilt the relevant overlap region is further extended by about 70~m. Currently the hole is being kept open by using an industrial anti-freeze agent, ESTISOL\textsuperscript{TM}-140, to allow for various scientific experiments~\cite{UV_logger, luminescence_logger}. 

\subsection{The Measurement Method}
\label{method}
\indent The measurement principle is illustrated in Figure~\ref{fig:diagram}. When a light source at the bottom of the device illuminates the ice, a fraction of the back-scattered light is detected by the camera. Data are collected at different depths and orientations. Measurements of the scattering length will be based on only the shape of the back-scattered light in the images as the cameras and LEDs are not absolutely calibrated. \\
\indent Internal reflections can contribute to background. To block the light reflected at the surface of the hole, a ring of brush is attached to the middle flange. Similarly, the light reflected at the glass is effectively blocked by the tight holding structure and inner part of the middle flange. Lab measurements confirmed that the cameras do not detect any significant amount of light from the light source through internal reflections.  \\  
\indent During the deployment, the depth of the instrument in the hole is recorded at the surface of the ice with time stamps. Furthermore, the instrument writes a log of the image capture times so that the depth corresponding to each image can be determined later. The orientation of the vessel changes slowly in the hole, due to the tension of the winch cable that holds the vessel. Although the rotation cannot be controlled, the orientation of the instrument is logged continuously using a 3-axes magnetic field sensor along with time stamps. This way the orientation of each image can also be determined. \\
\indent When the cameras capture images, their image sensors are exposed for 10~ms to 6~s depending on settings, and photons are collected over a range of orientations and depths. To precisely determine the orientation and depth corresponding to each image, exposure times are limited. In order to detect sufficient light even with short exposure times, it is crucial to use an intense and focused beam of light. \\
\indent Although it is possible to use only one pair of cameras and light sources for the planned measurement, three pairs are used in the same system to increase the total data rate. While one camera saves an image to its FLASH memory, a different camera can run with only the corresponding light source turned on.    

\section{The Prototype of the SPICEcore Hole Camera System}

\begin{figure}
\centering
\includegraphics[width=.0435\linewidth]{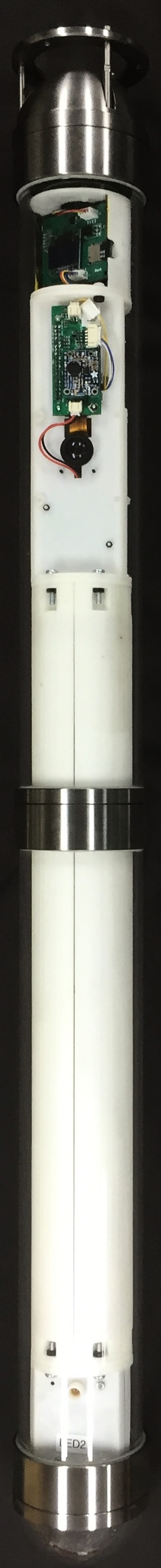}
\includegraphics[width=.3\linewidth]{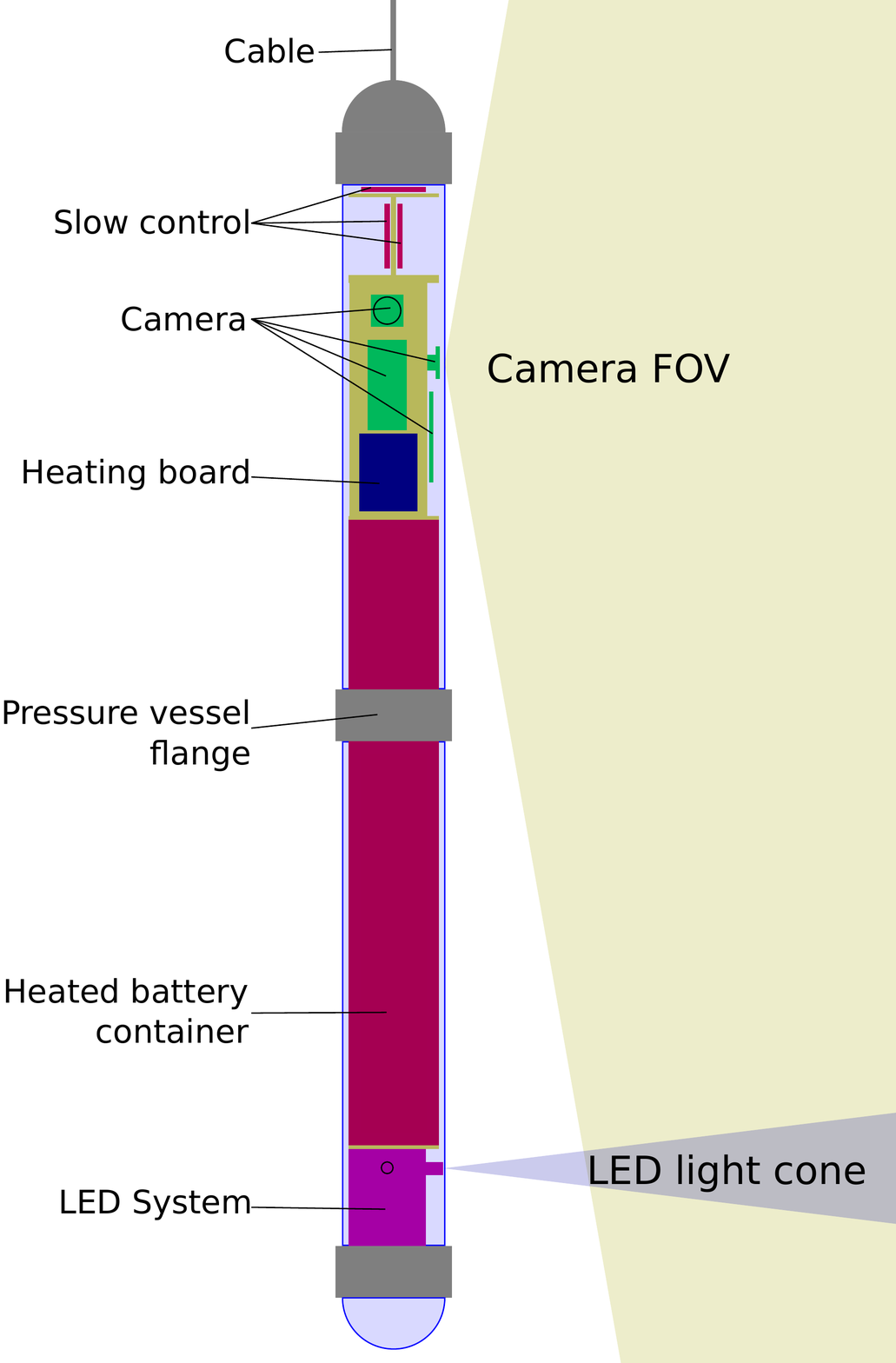}
\caption[margin=1cm]{{\bf Left:} A photo of the assembled system, {\bf Right:} A diagram of the SPICEcore hole camera system and its measurement principle. }
\label{fig:diagram}
\end{figure}
The prototype of the SKKU SPICEcore hole camera system was developed and assembled in 2017--2018 and shipped to the South Pole for the first deployment. Figure~\ref{fig:diagram} shows a photo of the prototype. The instrument consists of five main components: pressure vessel, camera module, light source module, battery, and heater.    \\
\indent The pressure-resistant vessel consists of two Borosilicate glass cylinders separated by a titanium flange. The cylinders are 40~cm long and 5.1~mm thick, and the material was chosen to have a high transparency to visible light. The top-end cap which connects the entire system to the winch cable is made of titanium, while a glass hemisphere is attached to the bottom of the cylinder through a titanium flange. Titanium was favored over other metals, as its thermal expansion coefficient is similar to that of the glass. The vessel was produced by Nautilus Marine Service GmbH, and its depth rating is 2,600~m in water. \\
\indent The camera module contains three identical cameras with SONY IMX219 CMOS color image sensors. Each camera is capable of capturing high resolution (up to 8 Megapixels) images and storing the data on a 256~GB microSD card. These cameras are equipped with fish-eye lenses and oriented to face outwards with 120 degrees of angular separation from each other. The Field-of-View of the cameras in ice is calculated to be 89$^{\circ}$ in horizontal direction and 73$^{\circ}$ in vertical direction, taking into account the refractive indices of air, Borosilicate glass, and ice. \\
\indent In the light source module, there are three LEDs each vertically aligned with one of the cameras. The LEDs emit narrow and intense beams of light. Their full width at half maximum (FWHM) is $7^{\circ}$, and the output power of the light flux is 170~mW under nominal operation conditions. The central wavelength of the LED light is 470~nm. This wavelength was favored, because the IMX219 sensor is particularly sensitive to this light, and it has been reported that the scattering in the ice is stronger with shorter wavelengths~\cite{SPICE}. \\
\indent In the slow control system, an ATmega2560 8-bit Micro-controller Unit (MCU) from Microchip is responsible for monitoring and logging environmental variables. The orientation of the instrument is measured using a PNI RM3100 3-axes magnetic field sensor. Temperatures at the light source module, battery, and the camera module are measured using TI TMP75C sensors. The MCU also communicates with each of the cameras via three UART channels. One of the reasons for choosing this MCU was that existing C libraries for the Arduino platform can be used in the development of the firmware. \\
\indent The system is powered by batteries which contain rechargeable lithium-ion cells. A dedicated heating system is installed to keep the battery temperature well within their operating temperature range, even if the ambient temperature can reach a low of $-50^{\circ}C$. By monitoring the battery temperature and controlling the heating power with the MCU, the batteries can run for more than 7 hours, which is long enough for performing measurements while in the hole. In this version of the device, one battery was dedicated to the heater, while the other two supplied power to the rest of the system.\\
\indent The prototype is fully autonomous and does not require any external power supply nor data interface. Operational modes can be selected via an IR remote controller at the surface immediately prior to deployment.

\section{The First Measurement at the South Pole}
\subsection{Deployment Summary}
\begin{wrapfigure}{R}{0.29\linewidth}
\centering
\includegraphics[width=1\linewidth]{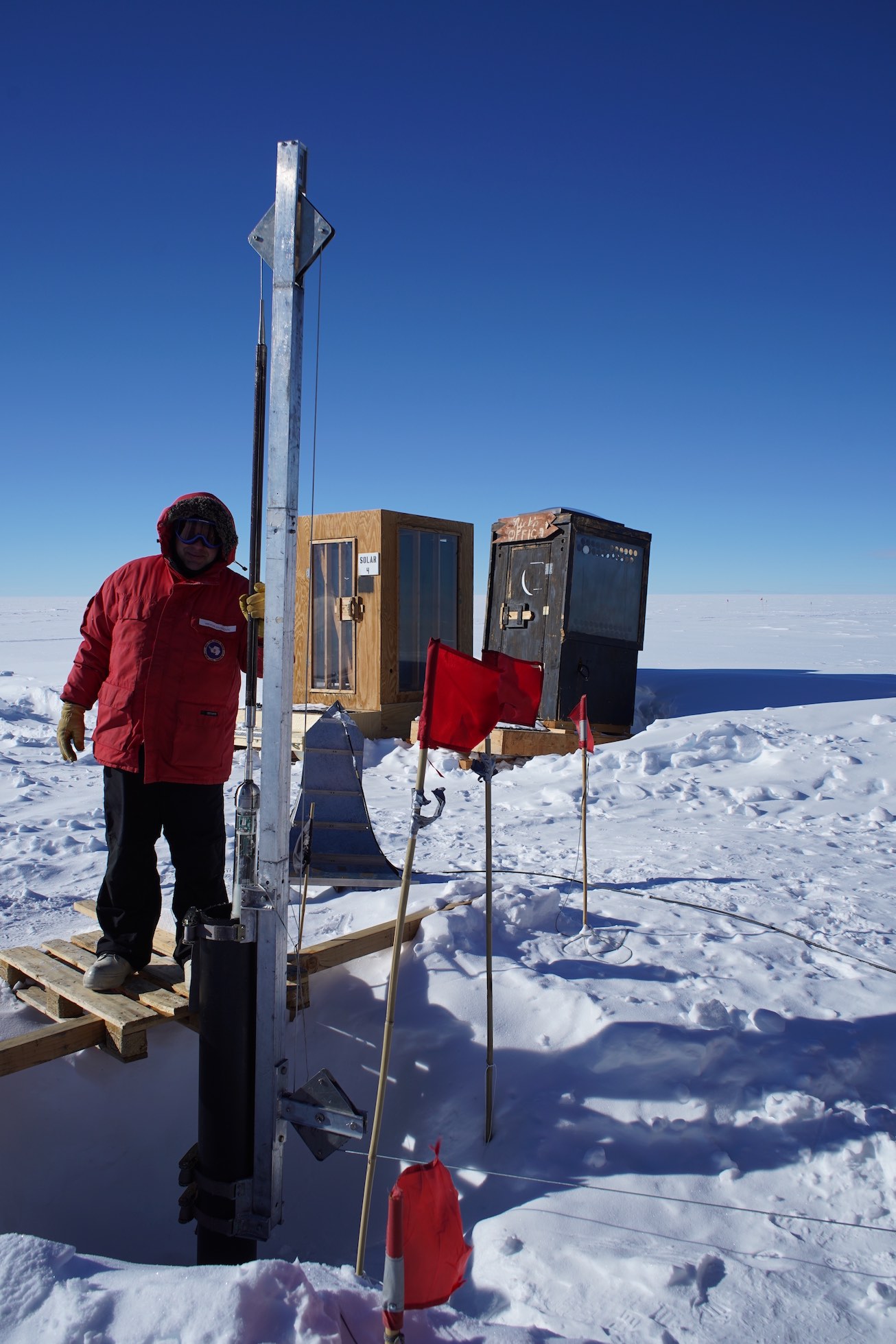}
\caption[margin=1cm]{A photo of the device taken at the SPICEcore hole immediately prior to deployment.}
\label{fig:deployment}
\end{wrapfigure}

The camera system was deployed at the South Pole during the 2018/2019 austral summer season. The vessel was lowered down to the depth of 1,695~m. The descent speed was kept roughly constant at around 10 m/min. The vessel was left at 1,695~m depth for 1.5~h and then raised up again at the same speed as the descent. The total deployment time was around 7.5~h. A total of 413 images were taken during the deployment. The exposure time values used for the images range between 10~ms and 6~s. A photo taken at the hole before the deployment is shown in Figure~\ref{fig:deployment}.

\subsection{Results}
\begin{figure}
    \centering
    \includegraphics[width=\linewidth]{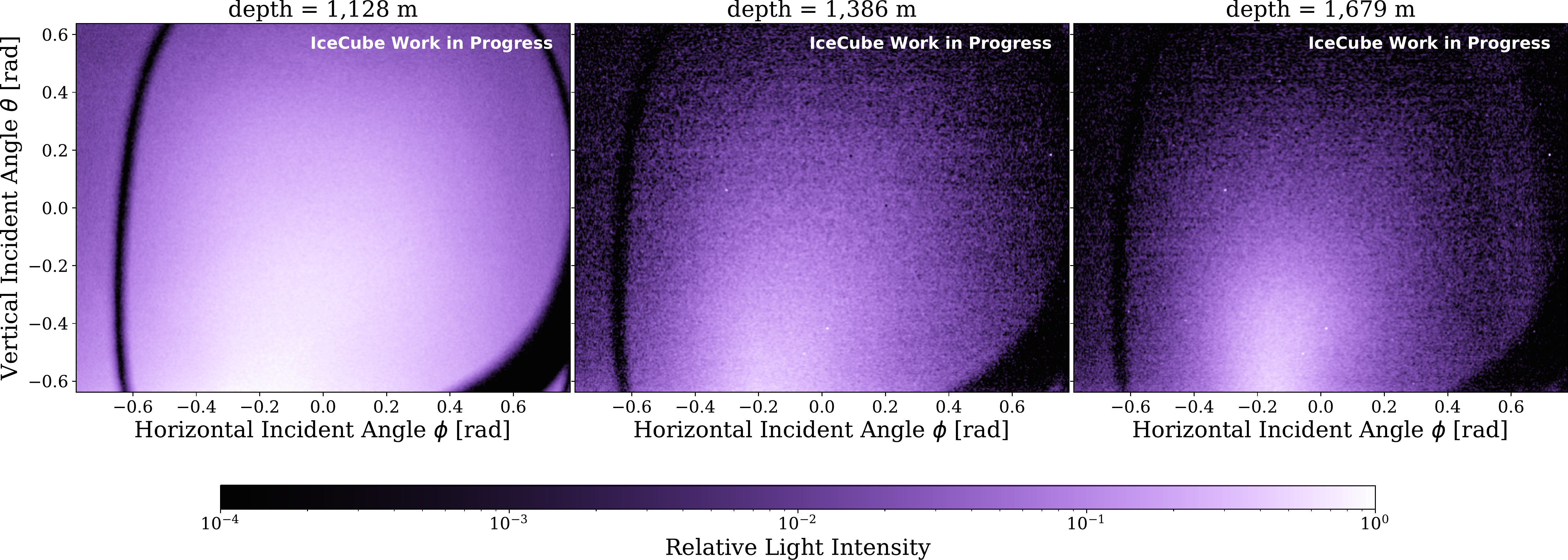}
    \caption[margin=1cm]{Images taken at different depths using one of the three cameras. The light incident angles are calculated based on the positions of pixels in each image and the Field-of-View of the camera. The dark areas that cut sharply into the light distribution stem from a harness that was used to hold the device. The light intensity in each image was scaled so that the maximum intensity is 1. The camera settings for capturing these images were the same.}  
    \label{fig:images}
\end{figure}

\begin{figure}[h]
\centering
\includegraphics[width=.55\linewidth]{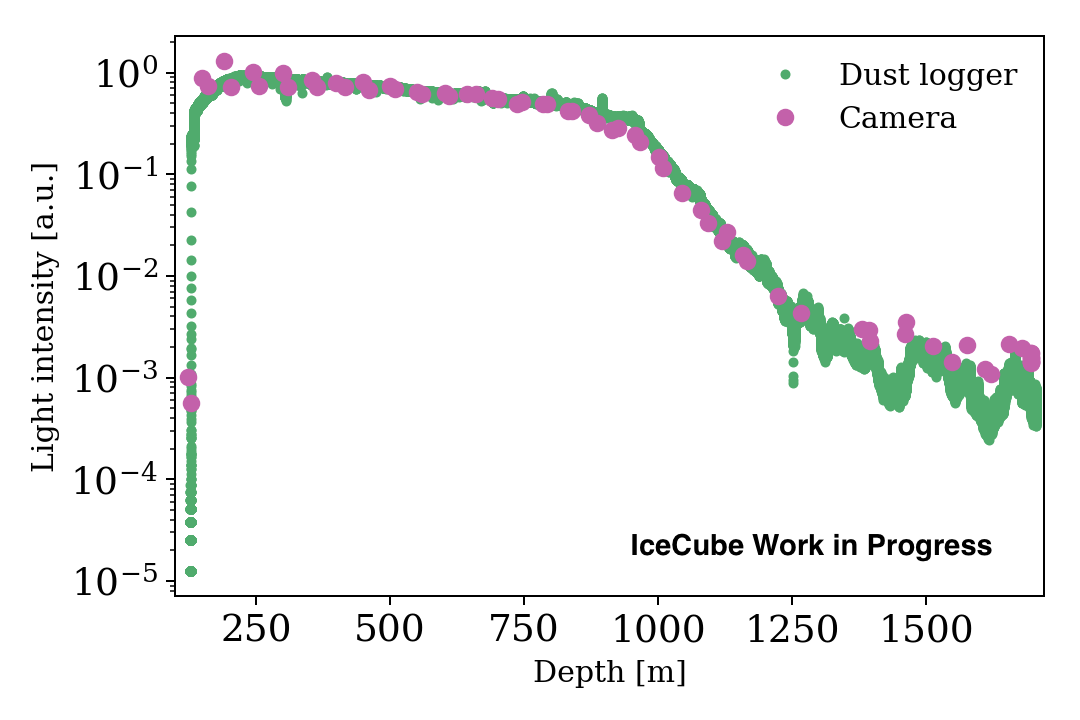}
\caption[margin=1cm]{The intensity of back-scattered light measured by the cameras and the dust logger is shown. The light intensity for the cameras is computed as the average pixel brightness, scaled to 1~s exposure time and shown as a function of depth. The dust logger points are scaled to match the data from the camera. }
\label{fig:dust}
\end{figure}

\begin{figure}
    \centering
    \includegraphics[width=\linewidth]{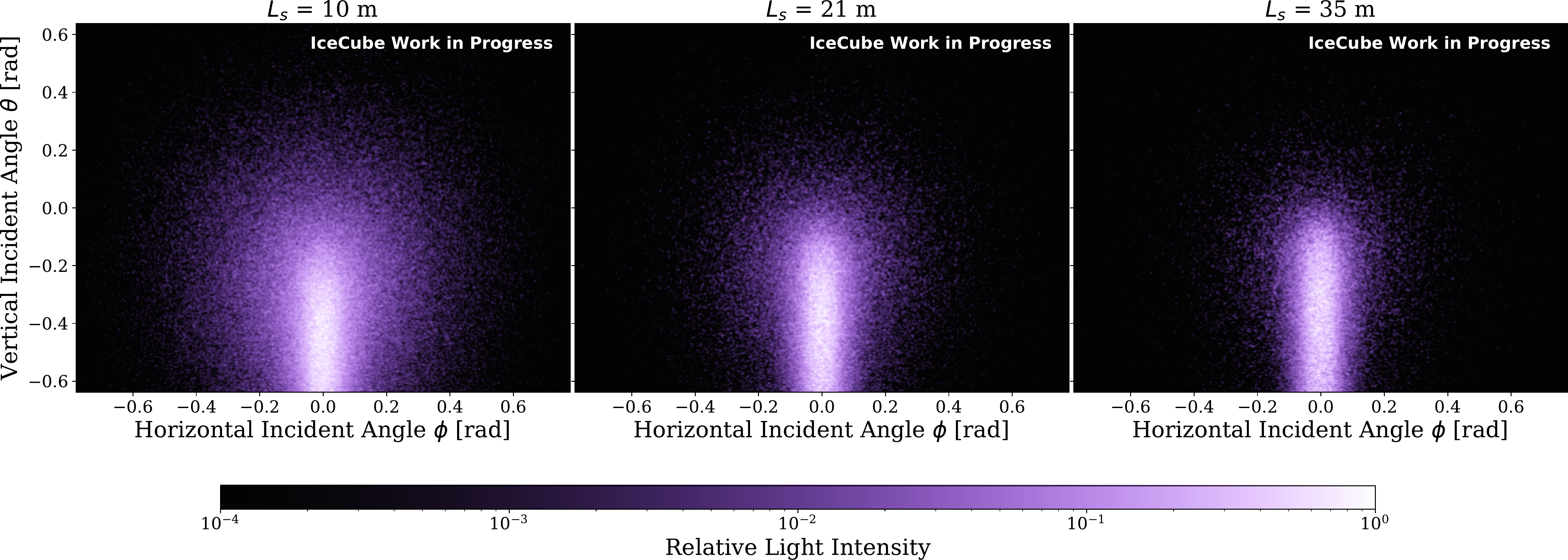}
        \caption[margin=1cm]{Images simulated using photon propagation software~\cite{PPC}. These images represent different effective scattering lengths ($L_{s}$). The light intensity in each image was scaled so that the maximum intensity is 1.}    
    \label{fig:MC}
\end{figure}

\begin{figure}[h]
\centering
\includegraphics[width=.55\linewidth]{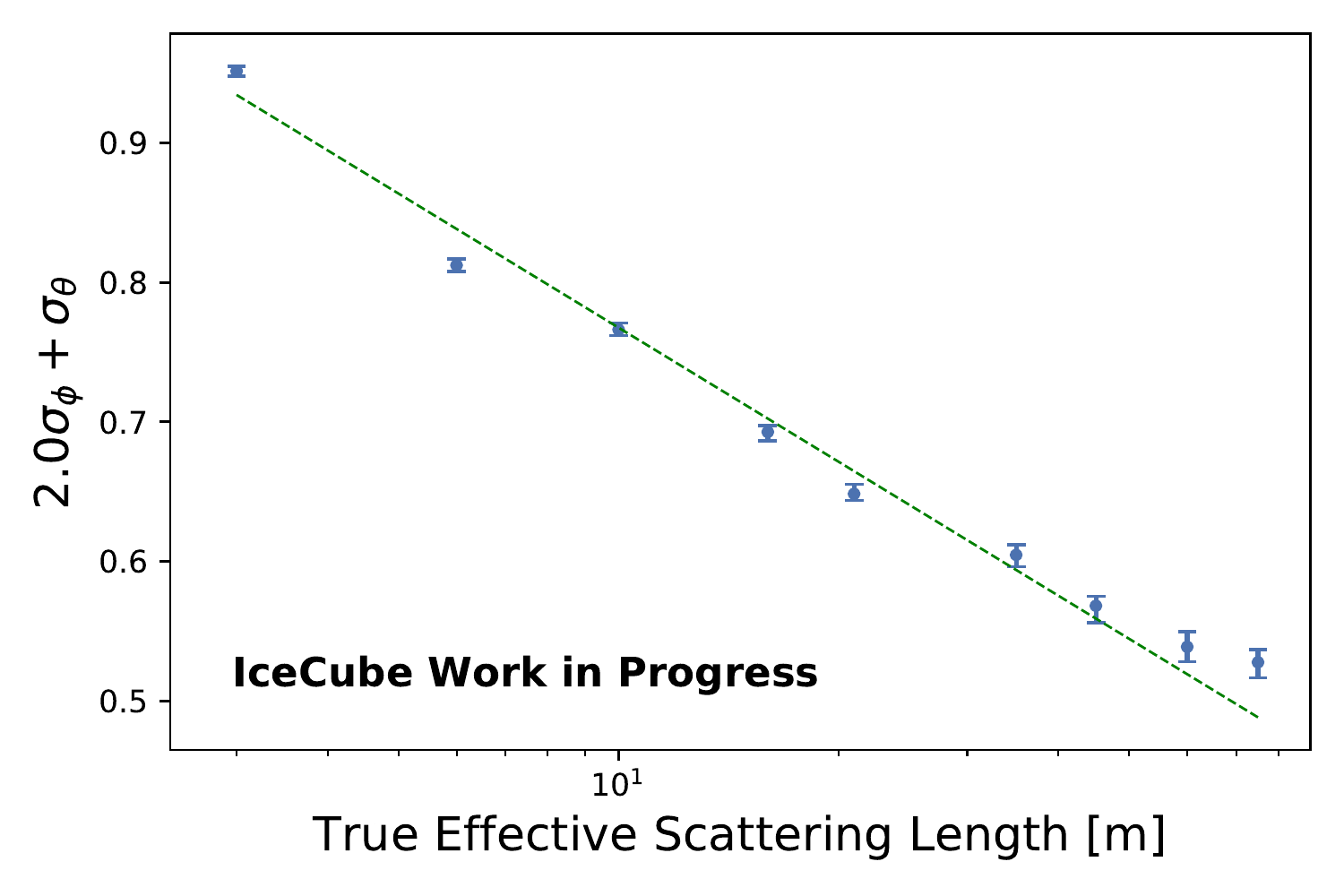}
\caption[margin=1cm]{The spread of the parameter $2.0\sigma_{\phi}+\sigma_{\theta}$ is shown for the nine Monte Carlo data sets corresponding to different true geometric scattering lengths. This linear combination is chosen to better differentiate different effective scattering lengths. The error bars represent 68\% confidence intervals.}
\label{fig:corr}
\end{figure}

Images taken at different depths show differences in brightness and light  distribution. In Figure~\ref{fig:images} sample images taken at 1,128~m, 1,396~m, and 1,679~m are shown to demonstrate these differences. In these images the measured light intensity is shown as a function of the horizontal and vertical angle $\phi$ and $\theta$ of incident light. With increasing depth the ice becomes more clear. This reduces the size of the illuminated area in the picture.\\
\indent In Figure~\ref{fig:dust} the intensity of the light measured with the cameras with an exposure time of 0.1s or longer are shown in the same graphic together with the intensity of the light independently measured with the laser-based dust logger~\cite{dust_logger}. The data were scaled to each other in order to illustrate the correlation between these measurements. The onset of the transition region between the bubbly ice and the clear ice is clearly visible at around 1,000~m.\\
\indent Images were simulated using photon propagation software~\cite{PPC} for different values of the scattering length of the light in ice. Images simulated for three different effective scattering lengths are shown in Figure~\ref{fig:MC}. As can be seen the distribution of the simulated photons is similar to the illuminated area shown in Figure~\ref{fig:images}. The simulation takes into account the angular emission profile of the OD-469L LED measured in the SKKU lab. Systematic effects, such as the alignment between the cameras and the LEDs, and the position of the harness, explain the differences between simulated images and the recorded images.\\
\indent To show the correlation between the shape of the scattered light in the image and the scattering length two width parameters $\sigma_\phi$ and $\sigma_\theta$ were defined. These are the root mean square of the light intensity distribution projected on the $\phi$ and $\theta$ axis respectively. These parameters are calculated using the equations

\begin{equation}
    \sigma_\theta = \sqrt{\sum_{i=1}^{N_{sim}} \frac{(\hat{\theta} - \theta_i)^2}{N_{sim}}}
\end{equation}

and

\begin{equation}
    \sigma_\phi = \sqrt{\sum_{i=1}^{N_{sim}} \frac{(\hat{\phi} - \phi_i)^2}{N_{sim}}},
\end{equation}

where $\theta_i$ and $\phi_i$ are the vertical and horizontal arrival angle of the photon number i. $\hat{\theta}$ and $\hat{\phi}$ are the average horizontal and vertical arrival angle of simulated photons. $N_{sim}$ is the total number of simulated photons.\\
\indent The combination $2.0\cdot\sigma_\phi +\sigma_\theta $ has then been calculated as a parameter that shows a particularly strong correlation to the true scattering length in simulations. This correlation is shown in Figure~\ref{fig:corr}.\\
\indent Using this parameter to compare simulations for different values of the scattering length to captured images, an analysis method to estimate the scattering length of light in the Antarctic ice is being developed. Before this method can yield reliable results a variety of systematic effects, such as the shadow of the harness around the pressure vessel and the effect of the ESTISOL\textsuperscript{TM}-140, have to be studied.\\

\section{Conclusions}

We have developed a camera system contained in a vertical pressure vessel suited to deployment in drilled holes in ice to measure the optical properties of the surrounding ice. The system was constructed and tested at the SKKU lab and deployed in the SPICEcore hole at the geographic South Pole during the 2018/2019 austral summer season.\\   
\indent Precise understanding of the properties of the ultra-pure deep ice below the surface of the geographic South Pole is one of the main prerequisites for successful data analysis with the IceCube Neutrino Telescope. Our SPICEcore hole camera measurements are targeted at more precisely characterizing the ice properties near the IceCube detector site. The IceCube collaboration has reported evidence for an anisotropy in the light propagation in the ice~\cite{anisotropy}, but the exact mechanism and description are still under active investigation. In the future we expect to deploy an improved version of the camera system to measure direction dependence in the light propagation in the ice. The results obtained from the first deployment also serve as a proof of concept of a camera system developed for IceCube Upgrade~\cite{ICU_camera}.

\section*{Acknowledgement}
The authors would like to thank the SPICEcore collaboration for providing the borehole, the US Ice Drilling Program, the Antarctic Support Contractor and the NSF National Science Foundation
for providing the equipment to perform the described measurement and for their support at South Pole.

\end{document}